**Title of the article:**
A survey study on major technical barriers affecting the decision to adopt cloud services

**Authors:**
Nattakarn Phaphoom,  Xiaofeng Wang, Sarah Samuel, Sven Helmer, Pekka Abrahamsson

**Notes:**
- This is the author's version of the work.
- The definite version was published in: Phaphoom, N., Wang, X., Samuel, S., Helmer, s., & Abrahamsson, P. (2015) A survey study on major technical barriers affecting the decision to adopt cloud services. The Journal of Systems and Software 103 (2015) 167–181

**Copyright holder's version**
- Copyright owner's version can be accessed at https://doi.org/10.1016/j.jss.2015.02.002

# A survey study on major technical barriers affecting the decision to adopt cloud services

Nattakarn Phaphoom, Xiaofeng Wang, Sarah Samuel, Sven Helmer, Pekka Abrahamsson

In the context of cloud computing, risks associated with underlying technologies, risks involving service models and outsourcing, and enterprise readiness have been recognized as potential barriers for the adoption. To accelerate cloud adoption, the concrete barriers negatively influencing the adoption decision need to be identified. Our study aims at understanding the impact of technical and security-related barriers on the organizational decision to adopt the cloud. We analyzed data collected through a web survey of 352 individuals working for enterprises consisting of decision makers as well as employees from other levels within an organization. The comparison of adopter and non-adopter sample reveals three potential adoption inhibitor, security, data privacy, and portability. The result from our logistic regression analysis confirms the criticality of the security concern, which results in an up to 26-fold increase in the non-adoption likelihood. Our study underlines the importance of the technical and security perspectives for research investigating the adoption

## 1. Introduction

Cloud computing is considered by many a paradigm shift in computing, representing a fundamental change in the way IT services are developed, offered, acquired, and paid for (Marston et al., 2011; Voas and Zhang, 2009). Ideally, in this paradigm a cloud service provider owns and manages a pool of computing resources, serving the general public by means of multi-tenancy. Computing services can be acquired over the Internet through self-service interfaces and service usages are automatically metered, allowing a pay-as-you-go (PAYG) model (Armbrust et al., 2010). Currently, cloud services are available in at least three different models (Mell and Grance, 2009): Infrastructure-as-a-Service (IaaS), Platform-as-a-Service (PaaS), and Software-as-a-Service (SaaS). The services are usually deployed in one of the four settings, public, private, hybrid, or community cloud infrastructures.

The cloud promises tremendous benefits for enterprises (Khajeh-Hosseini et al., 2011; Phaphoom et al., 2013) such as financial efficiency (Armbrust et al., 2010; Marston et al., 2011; Rafique et al., 2011), operational excellence (Goodburn and Hill, 2010; Harris et al., 2010), and continuous innovation (Harris et al., 2010; Jasti et al., 2011; Mladenow et al., 2012). According to the Gartner hype cycle, cloud computing has progressed through a period of inflated expectations and has been predicted to reach mainstream adoption in 2–5 years (LeHong and Fenn, 2013). The annual survey conducted by Northbridge and Gigaom (North Bridge Venture Partners, 2013), based on 855 respondents, reveals continuous growth of cloud demand. From 2012 to 2013 IaaS's growth rate was at 29%, followed by PaaS (22%) and SaaS (15%). A study by IDC on behalf of DG Connect of the European Commission (Bradshaw et al., 2012) has predicted that the European public cloud market will reach 11 billion Euros in revenue in 2014, increasing from 4.6 billion Euros in 2011. A global cloud provider survey conducted by KPMG (KPMG International, 2012b) also reports providers' expectations to increase their cloud revenue, in proportion to the total income, from 27% to 50% on average by 2014.

Although cloud computing promises great benefits to enterprise IT, we are still a long way off from treating computing resources like commodities such as electricity. Organizations are still reluctant to migrate their IT infrastructure into the cloud, fearing to lose control and other undesired outcomes.

Frameworks have been proposed to support the decision-making process for cloud adoption, considering various aspects of requirements, such as business objectives, demand behaviors, cost, technical and quality of services. Examples include a decision model supporting the selection of a cloud service provider based primarily on six requirement dimensions, namely flexibility, costs, IT security and compliance, scope and performance, reliability and trustworthiness, and service and cloud management (Repschlaeger et al., 2013). Focusing on service characteristics, a list of essential and high-level features of cloud services were discussed in several studies (Alhamad et al., 2010; Repschlaeger et al., 2012; Tamburri and Lago, 2011). Other work proposed methods for evaluating quality of services (Garg et al., 2011; Lee et al., 2009), and cost modeling (Khajeh-Hosseini et al., 2012; Klems et al., 2009; Mach and Schikuta, 2011). Based on such requirements with the goal of reducing barrier-to-entry, concrete factors that could potentially inhibit cloud adoption need to be identified.

Risks associated with underlying technologies, risks involving service outsourcing and enterprise readiness introduce significant barriers for the adoption of the cloud (Grobauer et al., 2011; Khajeh-Hosseini et al., 2011; Morgan and Conboy, 2013; Spring, 2011a, 2011b; Trivedi, 2013). Security and data privacy have been recognized as the most important barriers (Colt Technology Services Group Limited, 2011; Everest Global, 2013; IDG Enterprise, 2013; Lampe et al., 2013). Other obstacles consistently perceived by business executives, IT decision makers and business users are reliability and performance (Colt Technology Services Group Limited, 2011; North Bridge Venture Partners, 2013), integration challenges (Coopers, 2011; Everest Global, 2013; IDG Enterprise, 2013; KPMG International, 2012a, 2012b), and the complexity of existing IT infrastructure (NTT Europe Ltd., 2013). In terms of financial barriers, a lack of enterprise budget to invest in new initiatives (Everest Global, 2013), high implementation costs (KPMG International, 2012b) and difficulties in justifying the real value (Bradshaw et al., 2012; KPMG International, 2012a) have been identified. There are also several legal and organizational issues including data jurisdiction (Bradshaw et al., 2012), enterprise compliance (IDG Enterprise, 2013; KPMG International, 2012a) and insufficient knowledge (TechSoup Global, 2012). In order to address all these fears with the goal of reducing the barriers to entry, a first important step is to identify the concrete factors influencing the decision whether to move into the cloud or not.

Our study aims at understanding the impact of certain barriers (mainly technical and security-related) on the decision to adopt cloud services in an organizational context. In particular, we are interested in identifying the barriers that reduce the adoption likelihood. Top barriers recognized by industry are examined, namely availability, portability, integration with current enterprise systems, migration complexity, data privacy, and security. The concern over such barriers have been repeatedly reported since cloud computing was considered a hype (Colt Technology Services Group Limited, 2011; Coopers, 2011), and until recently (Everest Global, 2013; IDG Enterprise, 2013; NTT Europe Ltd., 2013). This probably suggests a lack of systematic and broadly informed approaches to overcome them. These barriers are also relevant to three general cloud service models. To reduce barrier-to-entry, their impacts on a decision of cloud adoption deserve further investigation.

We conducted a web-based survey to find out how individuals judge the importance of each barrier on the decision to adopt cloud computing. Based on this data we performed a statistical analysis to answer the following research question: *to what extent are the concerns over major barriers relevant to the decision of cloud adoption?* The findings suggest that not all the barriers significantly impacted the adoption decision. Instead, three potential adoption inhibitors were identified on the basis of group comparison, including portability, data privacy, and security. In particular, non-adoption likelihood may increase by a magnitude of 26, when a provider fails to demonstrate its adequate security safeguard.

The reminder of the paper is organized as follows. The next section reviews existing work on the perceptions of cloud barriers and identifies a set of common major barriers among the reviewed studies. It is followed by a discussion on the role and impact of these barriers on adopting organizations. Section 4 explains our research hypothesis and research framework, while the research methodology is described in detail in Section 5. Section 6 presents the results from a three-phase analysis, followed by a discussion on the findings. The paper concludes with implications for the cloud computing industry and directions for future work.

## 2. Related work

We divide the related work into two major parts: an overview of (industry) surveys discussing obstacles on the way to cloud computing adoption and a summary of analytical research investigating the factors influencing this decision.

### 2.1. Industry surveys on cloud adoption barriers

A number of cloud computing surveys have been conducted in industry. Many of their findings have been published and largely discussed among technology leaders and cloud specialists in technology management forums. We identified 11 publicly available studies conducted between 2011 and 2013 that specifically discuss barriers to cloud adoption.

Table 1 summarizes the characteristics of the identified industrial studies. The surveys were mostly commissioned by consultancy and IT service companies. They involve different groups of respondents, geography, data collection periods, and data collection methods:

- Six studies focus on collecting the views of business executives or IT decision makers. One specifically focuses on cloud providers and four capture the views of different players in the cloud ecosystem.
- Six studies are based on the data collected worldwide, two utilize data collected in Europe, and two studies focus on a specific country. Finally, one study does not clearly discuss the demography.
- In terms of the time period, five studies were conducted in 2013 and three each in 2012 and 2011.
- The data collection methods are online questionnaires or interviews (two studies apply both techniques). One study does not clearly specify its survey methodology.

After this brief overview, we now look at the individual studies in more detail.

The NTT survey (NTT Europe Ltd., 2013) was conducted to investigate the state of cloud adoption and to identify main adoption barriers for large organizations. Seventy-seven percent of the organizations cited in the report have adopted at least one type of cloud service to their ICT infrastructure. Fifty-six percent of the CIOs viewed the complexity of their IT infrastructure as the biggest barrier to large-scale adoption. For the public sector, the security issue was raised as the primary concern (89%).

A joint survey conducted by Everest Group and Cloud Connect (2013) aimed at identifying cloud adoption patterns, barriers, and decision making criteria. Security and integration challenges were mentioned as the most significant barriers to adoption from the view of consumers, providers, and advisors. Other top concerns from the consumer perspective were lack of budget for new initiatives, fear of vendor lock-in, and lack of suitable cloud solutions.

The IDC and TELUS study (Schrutt, 2013) focused on the Canadian cloud market. The study reports a great divide between perceptions of users and non-users of the public cloud. Seventy-six percent of the respondents with experiences in cloud computing perceived that the cloud is easy to use and integrate with other technology. In terms of security, a similar study conducted a year earlier identified difficulties in managing compliance and security as the main barriers. However, the more recent findings show that the cloud was perceived as a mechanism to improve data compliance (59%) and security (56%).

**Table 1**
Industrial cloud computing survey studies

| Id | Commissioned by (conducted by) | Data collection Year | Method | Number of respondents | Respondent profile | Geography |
|---|---|---|---|---|---|---|
| 1 | NTT *Vanson Bourne* | 2013 | Online questionnaire and interview | 300 | CIOs and senior IT decision makers in organizations with more than 250 employees | UK |
| 2 | Cloud Connect and Everest Group | 2013 | Online questionnaire | 302 | Cloud consumers, providers and advisors | EU |
| 3 | IDC and TELUS | 2013 | Telephone interview | 250 | Senior business and IT leaders from large organizations | Canada |
| 4 | IDG | 2013 | Interview | 1358 | Respondents stratified across CIO, Computerworld, CSO, InfoWorld, IT World and Network World websites | Global |
| 5 | North Bridge and GigaOM | 2013 | n/a | 855 | Cloud vendors, business users and IT decision makers | Global |
| 6 | KPMG *(Forbes Insights)* | 2012 | Online questionnaire | 650 | Senior executives at organizations using cloud | Global |
| 7 | KPMG *(Forbes Insights)* | 2012 | Online questionnaire | 179 | Cloud service providers | Global |
| 8 | TechSoup Global | 2012 | Online questionnaire | 10, 500 | Individuals who worked for an NGO | Global |
| 9 | IDC | 2011 | Interview | 1056 | IT decision makers | EU |
| 10 | PWC *(Bloomberg Businessweek)* | 2011 | Online questionnaire & interview | 489 | Business executives | Global |
| 11 | Colt *(Loudhouse)* | 2011 | Interview | 500 | IT decision makers in companies with a turnover of at least 100K Euro | EU |

The study by IDG (IDG Enterprise, 2013) shows that IT leaders continue to have concerns about security issues (66%), integration (47%) and an ability of the cloud to meet the standards of enterprises and the industry in general (35%). Another surveyed group, business leaders, had less concerns about integration but was more worried about the difficulty to measure return of investment (ROI).

The NorthBridge annual survey (North Bridge Venture Partners, 2013) captures the perceptions of cloud participants on drivers, inhibitors, and growth factors of the cloud market. Top inhibitors were found to be factors related to reliability, bandwidth, and complexity (65%), followed by factors related to regulation, compliance, and privacy (63%). The report conjectures that these are reasons for an increase in the adoption of private and hybrid clouds.

The two KPMG studies (KPMG International, 2012a, 2012b) we have identified focus on understanding cloud adoption patterns, perceived barriers, and the decision making process. The first study (KPMG International, 2012a) addresses the perceptions of cloud providers, whose biggest challenge was to demonstrate clear evidence of cost savings (38%). Moreover, they struggled to find strong business cases and pricing plans justifying a switch to cloud computing. Forty-eight percent of the providers believed that their clients' top concern was loss of control. This is followed by difficulties integrating existing architecture (41%) and data loss/privacy risks (39%).

The second study by KPMG (KPMG International, 2012b) addresses the views of cloud consumers. The findings indicate a shift of organizational focus away from a pure cost reduction objective toward the transformational aspects of adopting the cloud. The top challenges identified were the high costs of implementation, transition, or integration (33%), integration with existing infrastructure (31%), and data loss and privacy risks (30%).

The study by Techsoup (TechSoup Global, 2012) explores the usage of clouds by non-governmental organizations (NGO) in order to better understand benefits, costs, and barriers in this specific domain. The analysis is based on more than 10,000 responses. Lack of knowledge was perceived as the biggest barrier of cloud adoption, identified by 60% of the respondents. This finding was consistent across geographies and organizational sizes. Other top barriers were costs (49%), data security (45%), and lack of trust (45%).

The IDC study (Bradshaw et al., 2012) analyzes the demand for cloud computing in Europe and also potential barriers hindering its introduction. The most relevant factors found here include legal matters, security and data protection, trust, and difficulties in figuring out the cost/benefit ratio. The in-depth statistical analysis highlights the various degrees of impact each factor introduces in the short and long term, and in different organizational contexts. Furthermore, the study emphasizes the accumulative effect the barriers have on adopting the cloud.

The aim of the PwC study (Coopers, 2011) was to unveil the current state of data center management and the shift to cloud computing. Seventy-nine percent of the respondents had or were developing organizational cloud strategies. Regarding adoption barriers, security received the most attention (62%), although other issues appear to weigh heavily on executives' minds as well. Top concerns included data and system integration (42%), portability (41%), and viability of third-party providers (40%).

Finally, the study by Colt (Colt Technology Services Group Limited, 2011) addresses a wide range of risks. Nevertheless, security remains the most crucial barrier (63%), followed by integration (57%), and performance and reliability concerns (55%). Additionally, almost half (42%) of IT leaders believe that they are not in a position to be fully able to assess risks associated with cloud adoption. The findings further emphasize the need for a strategic approach to enterprise cloud adoption, rather than focusing merely on a technology-driven view.

We conclude by making the following observations. First of all, cloud surveys have been conducted periodically with an objective of capturing the adoption patterns, finding ways to penetrate the markets, and understanding factors that help in facilitating the adoption process. Secondly, the findings are backed with large numbers of responses and, in some cases, illustrated with the viewpoints of business or IT leaders of leading enterprises. Thirdly, in terms of analysis methods, most studies apply a frequency count to determine the top items. Only one study (Bradshaw et al., 2012) applies an in-depth statistical analysis to understand the significance of each factor.

We have also identified two important aspects concerning the main barriers. Firstly, security was consistently mentioned as a crucial factor. In four of the studies (Colt Technology Services Group Limited, 2011; Coopers, 2011; Everest Global, 2013; IDG Enterprise, 2013) security and integration issues were perceived to be the first and second most relevant barriers. Two of these studies were conducted in 2011, the others in 2013. This seems to suggest that a clear strategy to overcome such issues is still missing or has not been established or acknowledged by the community yet. Secondly, the main barriers vary depending on the area or domain. The study by Khajeh-Hosseini et al. (2011) provides a classification along the lines of technical, security, financial, organizational and legal issues. The classification

**Table 2**
Top barriers in different areas.

| Technical | Security |
|---|---|
| IT infrastructure complexity | General security concerns |
| Integration challenge | Data privacy |
| Vendor lock-in | Data loss |
| Reliability and performance | Lack of trust |

| Financial | Legal and organizational |
|---|---|
| No budget for new initiatives | Legal jurisdiction |
| High implementation cost | Enterprise compliance |
| Difficult to evaluate benefits | Insufficient knowledge |
| | No suitable solution |
| | Viability of third-party vendor |

of top barriers recognized through this review is briefly summarized in Table 2.

### 2.2. Determinants for cloud adoption

There has also been quantitative research on a more theoretical basis in the area of cloud adoption with the goal of identifying and measuring the influence of various determinants. We have summarized important studies and their key findings in Table 3.

There are two major frameworks or models that have been used extensively in IT and IS research to investigate innovation adoption: the Technology-Organization-Environment (TOE) framework proposed by Tornatzky et al. (1990) and the Diffusion of Innovation (DOI) theory proposed by Rogers and Rogers (2003). They complement each other, thus providing a more complete view on this matter (Oliveira, 2011). In fact, in a number of studies the five innovation characteristics suggested by DOI, relative advantages, compatibility, complexity, trialability, and observability, have been integrated into the TOE framework to further explain its technological context.

To some extent all three contexts of the TOE framework have been considered in Espadanal and Oliveira (2012); Hsu (2013); Lian et al. (2014); Lumsden and Gutierrez (2013); Tehrani (2013), with two studies also looking at the characteristics of decision makers (Lian et al., 2014; Tehrani, 2013). The *technological context* of the TOE framework refers to the technologies available for organizations and their characteristics that impact the adoption process. Along with the first three factors of DOI and other factors, such as security and cost, this is the context investigated in the studies published in Espadanal and Oliveira (2012); Lian et al. (2014); Tehrani (2013). The *organizational context* concerns the structure, processes, and resources of an organization that support or restrict the adoption of innovation. Various sets of factors were considered relevant, including top management support (Espadanal and Oliveira, 2012; Lian et al., 2014; Lumsden and Gutierrez, 2013), a firm's size (Espadanal and Oliveira, 2012; Lumsden and Gutierrez, 2013), technology readiness (Lumsden and Gutierrez, 2013), information intensity (Tehrani, 2013), employees' cloud knowledge (Tehrani, 2013), and IT capability (Hsu, 2013). In the *environmental context*, external factors, such as industry, competitors, laws and regulations, introduce opportunities or constraints for the adoption of cloud services.

**Table 3**
Quantitative studies investigating dominant factors of enterprise cloud adoption.

| Study and context | Framework | Analysis method | Valid responses | Key findings |
|---|---|---|---|---|
| (Lian et al., 2014) Hospital in Taiwan | 2 *Human* factors (CIO innovativeness, technical competence), 4 *technological* factors (security, complexity, compatibility and cost), 4 *organizational* factors (relative advantages, management support, resources and benefits) and 2 *environmental* factors (government policy and industry pressure) | Mean value and ANOVA | 60 | The most important context is technology, followed by human, organization and environment. Data security is perceived as the most critical concern. |
| (Lumsden and Gutierrez, 2013) UK | 3 *Technological* factors (relative advantage, complexity, compatibility), 3 *organizational* factors (management support, firm size, technology readiness) and 2 *environmental* factors (competitive pressure and trading partners) | Factor analysis | 257 | Compatibility, relative advantage, technology readiness and management support appear as the most important adoption determinant. |
| (Tehrani, 2013) SMEs | 2 *Human* factors (decision maker's innovativeness and cloud knowledge), 6 *technological* factors (relative advantage, complexity, compatibility, cost, security and trialability), 2 *organizational* factors (information intensity and employee's cloud knowledge) and 2 *environmental* factors (external support and competitive pressure) | Logistic regression | 101 | Decision maker's knowledge of the cloud is the only significant factor positively related to adoption decision. |
| (Hsu, 2013) Taiwan | 2 *Technological* factors (perceived benefits and business concerns), 1 *organizational* (IT capability) factor and 1 *environmental* factor (external pressure) | Structural model | 200 | Perceived benefits have the strongest influence on the adoption, while the external pressure is not a significant factor. |
| (Espadanal and Oliveira, 2012) Manufacturing and service sectors in Portugal | 6 *Technological* factors (relative advantage, complexity, compatibility, security and privacy, cost saving and technology readiness), 2 *organizational* factors (management support and firm size) and 2 *environmental* factors (competitive pressure and regulatory support) | Structural model | 249 | Management support, relative advantage, and technology readiness are the strongest adoption determinants. The environmental construct does not significantly impact the decision. |
| (Wu, 2011) SaaS | 7 Factors including security and trust, marketing effort, perceived benefits, attitude toward technology innovations, social influence, perceived usefulness, and perceived ease of use | Structural model | 42 | The factor with the highest explanation power is perceived usefulness, followed by perceived ease of use and attitude toward innovation. |
| (Gupta et al., 2013) SMBs in Asia-Pacific | 5 Factors including cost, ease of use, reliability, sharing and collaboration, and security and privacy | Structural model | 211 | Ease of use is the most favorable factor, leading to fast adoption. SMBs appear quite satisfied with the security and privacy of the cloud. |
| (Alexander and Thomas, 2011) SaaS in Germany | 5 Factors of *perceived risks* (performance, economics, strategic, security, and managerial risks) and 5 factors of *perceived opportunities* (cost, flexibility, focus on core competencies, access to specialized resources, and quality improvement) | Structural model | 349 | Perceived opportunities have a much stronger impact than perceived risks do. Security threats are the dominant factor for overall risk perceptions. |
| (Our current study) | 6 *Major technical barriers* (availability, portability, | Logistic regression | 352 | Security concern is the most critical |

| | | |
|---|---|---|
| Global cloud adoption | integration, migration complexity, data privacy, security) | inhibitor for cloud adoption. |

The TOE framework and DOI are clearly not the only models used in research, Wu employed an adapted version of the Technology Acceptance Model (TAM) to investigate SaaS adoption intentions (Wu, 2011). The original version of TAM explains the technology adoption by considering the perceived usefulness, perceived ease of use, and behavioral intention. The adapted model is integrated with several factors derived from Rogers' diffusion theory, such as media and social influence, flexibility, perceived status benefits, and attitude toward mobile innovation. Moreover, security and trust as well as marketing efforts were considered impactful for SaaS and consequently were also included in Wu's model.

Two studies focused particularly on the technological aspect of cloud services. Gupta et al. (2013) proposed a model which concentrates on the benefits of cloud services and its reliability and security characteristics. Alexander and Thomas (2011) employed an opportunity-risk model, achieving good results with it.

In the studies discussed above the technological aspect of cloud computing has been recognized as an influential adoption determinant. For instance, in healthcare, Lian found that among the four aspects considered, the technological angle is the most impactful one (Lian et al., 2014). The studies conducted in Taiwan (Hsu, 2013), the United Kingdom (Lumsden and Gutierrez, 2013), and Germany (Alexander and Thomas, 2011) identified the *perceived advantages* promised by cloud services as a significant adoption driver. Along these lines, the perceived usefulness (Wu, 2011) and ease of use (Gupta et al., 2013) were also recognized as the best explanatory factor for the adoption. Different results were obtained in some other contexts. For SMEs, Tehrani found the decision maker's knowledge of the cloud as a definite adoption determinant (Tehrani, 2013). Top management support was the most influential factor in the manufacturing and services sector (Espadanal and Oliveira, 2012).

Judging the different models in terms of their explanatory power, the opportunity-risk model (Alexander and Thomas, 2011) performs best, explaining 83% of the variance in the intention to increase SaaS adoption. The adjusted TAM model provides 52.6% prediction accuracy of the usage intention (Wu, 2011). The technological and TOE-based models, i.e., (Espadanal and Oliveira, 2012; Gupta et al., 2013), explain less than 50% of the adoption variance.

## 3. Identification of barriers to cloud service adoption

The study by Khajeh-Hosseini et al. (2011) identifies 39 risks associated with using a public IaaS cloud. With a focus on technical and security related aspects, we have chosen six of those to include in our survey, basically those applying to the three general cloud service models, and those that have been recognized by industry (Colt Technology Services Group Limited, 2011; Coopers, 2011) as major barriers to adoption. The six barriers represent an important aspect of the technical barriers in Table 2 and two security-related barriers, which can potentially be affected or partially be managed by technical means. Please note that Table 2 was created based on industrial surveys published in 2011–2013 and it was not our intention to make a direct one-to-one mapping from it to the six barriers chosen by us. Nonetheless many of the aspects in Table 2 can be found in the selection we made. For example availability is a crucial component of reliability; infrastructure complexity leads to migration challenges; vendor lock-in is also known as portability issue; integration challenges, security, and data privacy were directly considered. The concerns over such barriers have also been reported in more recent surveys, such as in Everest Global (2013); IDG Enterprise (2013); NTT Europe Ltd. (2013), which probably suggest that ways to overcome them are not yet largely informed. Therefore, to reduce barrier-to-entry, their impacts on a decision of cloud adoption deserve further investigation. In the following, we describe these factors in more detail.

### 3.1. Availability

Continuous availability and convenient access to business data and services are aspects of availability that are fundamental for the efficient execution of processes. For this reason, availability, which implies service uptime and reliable performance, is one of the most important quality measures of cloud-based services. However, the lack of availability or failure on the provider side of delivering the promised level of availability has caused hesitation among potential adopters. A content analysis over a major cloud computing discussion forum (Phaphoom et al., 2012) reveals constant concerns about incidents such as large-scale service outages (Miller, 2012) and the unexpected shutdown of virtual machines. Consequently, the concern over service availability may inhibit the adoption of cloud services.

### 3.2. Portability

Data lock-ifor Software-as-a-Service (SaaS) models and system lock-in for Infrastructure-as-a-Service (IaaS) and Platform-as-a-Service (PaaS) models are a potential issue, creating a dependency of a customer on the proprietary services and architectures of a specific provider (Dowell et al., 2011). Such a dependency considerably reduces the ability of a client to migrate cloud artifacts from one provider to another or move the artifacts back in house, or at least makes this an expensive endeavor. The lock-in issue turns critical when considering to terminate a contract with a particular provider. This factor will continue to be an inhibitor as long as there is a lack of standardization on service APIs, architectural components, and data structures (Armbrust et al., 2010; Rodero-Merino et al., 2010).

### 3.3. Integration with current enterprise systems

Certain organizational requirements, such as the need to integrate workflows that are partially executed in the cloud and partially on an in-house enterprise system, lead to the necessity to establish a hybrid cloud solution. Data integration is not straightforward and among the challenges are ensuring interoperability and maintaining uniform control and transparency over all resources. Integration at minimum requires stable connectivity between partners and a standardization of components or the integrated data. Poorly managed integration, where data is stored in different formats in multiple data stores, leads to so-called information silos (Curbera, 2013). So it is not surprising that integration was reported to be one of the most time consuming aspect of implementation on the client side (MuleSoft, 2011). Although approaches for integrating systems are available, e.g. Breiter and Naik have identified three integration patterns requiring different implementation strategies (Breiter and Naik, 2013), this still remains a barrier.

### 3.4. Migration complexity

Due to the complexity and risks associated, it is generally advised to migrate a non-trivial system to the cloud in several stages, starting with the less critical parts of the system. In this way, the true impacts of the cloud are gradually recognized and potential pitfalls can be mitigated as the adoption plan proceeds. In fact, cloud adoption programs take up to several years for large enterprises and governments (Trivedi, 2013). In the case of SaaS subscriptions, corporate customers can opt for trial periods or trial modules to examine the quality of service and understand the implications on business processes (Conboy and Morgan, 2012). An iterative migration is still far from trivial, as issues such as complicated software licensing schemes and lack of personnel and funding for supporting new initiatives or feasibility studies can impede the process. So the sheer complexity of the migration task may hold back an organization from adopting cloud services.

*3.5. Data privacy and legal concerns*

When the data and workload are stored and processed on-premise, an organization has ultimate control over the sensitive data throughout its lifetime. Once these responsibilities are outsourced to the cloud, an organization needs to verify that the cloud provider respects the regulatory and compliance requirements. This is especially true for critical domains such as government, finance, and healthcare. Organizations operating in these environments need to adhere to specific regulations, such as the Health Insurance Portability and Accountability Act (HIPAA), the Payment Card Industry Data Security Standards (PCI DSS), and the Federal Information Security Management Act (FISMA). General concerns in this context include the need for informed consent from users when dealing with personal data, the need for strong access mechanisms, compliance to data jurisdictions, and compliance to data confidentiality regulations (Cloud Security Alliance, 2011). So the possible lack of enforcement of data privacy regulations is yet another obstacle on the way to adopting cloud services.

*3.6. Security*

Security has always been an important factor in information systems and cloud services are no exception. Surveys on cloud security (Phaphoom et al., 2013; Subashini and Kavitha, 2011) have identified security issues associated with cloud service models, architectural layers, and the technology underlying the cloud. The Cloud Security Alliance has determined main threats, including insecure interfaces, malicious insiders, shared technology vulnerabilities, data loss and data leakage, and service hijacking (Cloud Security Alliance, 2010). Inappropriate mechanisms for physical access control, authentication, and authorization account for many of such incidents. In the public cloud, however, many of these critical mechanisms fall under the sole control of a provider (often this boils down to a matter of trust as well (Phaphoom et al., 2012)). Unsurprisingly, many potential customers are reluctant to give up this control.

**4. Research question and hypotheses**

Our research goal was to assess the relevance of the adoption barriers on an enterprise cloud adoption decision. We focused particularly on the major barriers identified on the basis of existing survey findings. Section 2.1 provides more details on the identification of such barriers. The general research question that guides the study was formulated as:

*RQ: to what extent are the concerns over major barriers relevant to the decision of cloud adoption?*

The research framework is represented in Fig. 1. It consists of one dependent variable, *i.e. the adoption decision*, six independent variables, *i.e. concerns over the barrier factors* and two control variables, *i.e. geographical location and business domain*. A set of barriers included in the study are service availability, portability, integration, migration complexity, data privacy and security. As discussed in Section 3, when not properly managed, such factors may lead to undesired outcomes for enterprises adopting the cloud. Consequently, to some extent they influence the adoption decision. The adoption patterns and tendency vary across regions and business domains (Coopers, 2011), therefore, the potential effects of contextual differences are controlled to investigate the relevance of the independent variables.

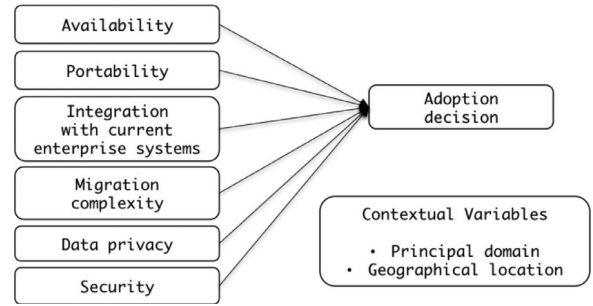

**Fig. 1.** The research framework.

We formulated the hypotheses on two levels, individual and integrated, to investigate the posed research question. The aim of the individual-level hypotheses was to assess the relationship between an individual independent variable and the outcome variable. This was done by investigating differences in distributions between a group of non-adopters and a group of adopters with regards to the influence of each barrier in the decision making process. The template hypothesis, in which *[the barrier]* is a placeholder for one of the concrete barriers identified earlier, is defined as follows:

*H_individual: There is a significant difference in the degree of concerns over [the barrier] between a group of organizations that adopt cloud computing and a group of organizations that do not adopt cloud computing.*

At the integrated-level, we evaluated the contribution of the entire research model in explaining the variation in the cloud adoption decision. It was observed in the study of IDC by Bradshaw et al. (2012) that the real obstacle hindering the growth of the cloud market might stem from the accumulative impact of barriers. Such a claim can be investigated through the hypothesis at the integrated-level, which was formulated as:

*H_integrated: Combined influences of different barriers can be used to determine cloud non-adopters from adopters.*

**5. Research method**

We have conducted a web-based survey to collect the empirical data and performed statistical analysis to examine the hypotheses. The design of the survey, data characteristic and analysis approach are explained in this section.

*5.1. Survey design*

The data was collected by means of a web-based questionnaire. The questionnaire contained four main parts. The first part was introductory, providing a definition and examples of cloud services. In the second part, the respondents were asked to provide background information about their organization, including the principal domain, the country they work in, and their role within the organization. The third part asked which decision an organization had made in terms of adopting cloud services, i.e., it was adopted or not adopted. The fourth part contained detailed questions to assess the respondents' view on the influence of the barriers and other factors on the cloud adoption decision. Finally, in an optional section the respondents were given the opportunity to make comments or suggestions. The participants were also not obligated to answer all the questions, in order to avoid forcing the respondents to answer questions when they were not sure about the involved concepts.

**Table 4**
Measurement items of the independent variables.

| Variables | Measurement items |
|---|---|
| Availability (AV) | 100% uptime is important for your decision to migrate to cloud computing. |
| Portability (PO) | It is important to avoid being tied to one vendor when making the decision to migrate to the cloud. |
| Integration with current systems (IN) | The ability to create compatible connections between in-house software and the cloud services affects my decision to adopt cloud computing. |
| Migration complexity (MI) | The need to migrate all services at once is a major factor to consider when migrating to cloud computing. |
| Data privacy (DP) | The need to keep sensitive data private plays a central role in my decision to adopt cloud computing. |
| Security (SE) | Sharing data security information with a cloud computing vendor is a barrier to my organization adopting cloud computing. |

In terms of measures, each research variable was operationalized to one construct, respectively. Table 4 summarizes the measurement items used for the independent variables. The following defines the scale used for data collection.

- For the question concerning the adoption decision, the respondents were asked if the organization had decided to go ahead and migrate to cloud computing. Answers to this yes-no question were treated as a dichotomous variable. The score 0 was assigned to "yes" and the score 1 was assigned to "no" (due to our focus on the circumstances in which cloud computing was not adopted).
- In order to measure the level of concern over cloud barriers, we asked the respondents to express *a level of agreement that a specific barrier plays an important role* during the decision making process, or *a level of impact a barrier introduces* during the decision-making process. Answers were collected using a five-point Likert scale, with one of the two sets of alternatives:
    - Answers were collected with the alternatives of "strongly disagree", "disagree", "neutral", "agree" and "strongly agree". The score 1 was assigned to "strongly disagree" and 5 was assigned to "strongly agree", respectively.
    - Answers were collected with the alternatives of "very low impact", "low impact", "neutral", "high impact" and "very high impact". The score 1 was assigned to "very low impact" and 5 was assigned to "very high impact", respectively.

### 5.2. Data collection

The survey instrument was developed and hosted on SurveyMonkey. The survey was expected to reach mass public, as we constantly posted it on Linkedin, Linkedin's cloud computing groups, Twitter, Facebook and cloud computing forums. Personal invitations were also sent out to target subjects using aforementioned social media channels. The data collection lasted for approximately 2 months, from April to May 2011, receiving 443 responses. Four hundred thirty-two responses have completely answered questions on background information and a cloud adoption decision, which are mandatory for us. In total there were 352 complete responses (79.46%). Table 5 summarizes the sample distribution based on adoption status, organization's principal domain, geographical location and respondents' role. Please note that, to increase the chance of getting more responses, we did not ask the respondents about the organization they work and the role in the same time, as it may cause reluctancy to respond the survey. The unit of analysis is therefore professional's opinion, rather than an organization.

The open-ended comment is another valuable source of information obtained from the survey. About 6% of the respondents provided additional explanations or rationales to justify their choice. Such information was about (1) motivation and cloud adoption conditions in their context, (2) other challenges faced, and (3) the trend and usage behavior. Relevant findings are discussed along with the statistical results in Section 7.

**Table 5**
Sample demographic, calculated based on 352 respondents who provide complete answers on the major barriers.

| Categories | Number of complete responses |
|---|---|
| Cloud computing adoption | |
| – Adopter | 202 (57.39%) |
| – Non-adopter | 150 (42.61%) |
| Principal domain | |
| – Information Technology | 143 (40.63%) |
| – Telecommunication | 47 (13.35%) |
| – Professional and business services | 30 (8.52%) |
| – Government and non-profit | 23 (6.53%) |
| – Other[a] | 22 (6.23%) |
| – Healthcare | 21 (5.97%) |
| – Education | 17 (4.83%) |
| – Financials | 16 (4.55%) |
| – Manufacturing | 16 (4.55%) |
| – Other services[b] | 7 (1.99%) |
| – Sales and marketing | 5 (1.42%) |
| – Agriculture, mining and food | 4 (1.14%) |
| – Aerospace | 1 (0.28%) |
| Continent | |
| – Europe | 140 (39.77%) |
| – North America | 132 (37.5%) |
| – Asia | 57 (16.19%) |
| – Australia | 12 (3.41%) |
| – Africa | 6 (1.70%) |
| – South America | 5 (1.42%) |
| Respondent's role | |
| – IT/software development | 85 (24.15%) |
| – CEO/CTO/VP | 54 (15.34%) |
| – Sales/business development | 48 (13.64%) |
| – Project management[a] | 31 (8.81%) |
| – Other | 23 (6.53%) |

| | | |
|---|---|---|
| – Marketing | 20 | (5.68%) |
| – Business strategy | 19 | (5.4%) |
| – Compliance (e.g. legal, quality) | 16 | (4.55%) |
| – Engineering | 13 | (3.69%) |
| – Operation and support | 13 | (3.69%) |
| – Research and development | 10 | (2.74%) |
| – Human resource | 9 | (2.56%) |
| – Finance and accounting | 5 | (1.42%) |
| – Design | 4 | (1.14%) |
| – Health | 2 | (0.57%) |

[a] The classification was identified as 'Other' by the respondent.
[b] Other services include transportation, entertainment and construction.

### 5.3. Analysis approach

The data analysis followed three stages, examining data distribution, bivariate analysis, and multivariate analysis. The first step aimed at comparing the degree of importance of the barriers as perceived by the respondents. Then, bivariate analysis was applied to explore the relationship between a single barrier and the adoption decision. In attempting to select a multivariate method, we were considering discriminant analysis, logistic regression, and structural equation modeling (SEM). Logistic regression is considered the most appropriate one based on the design of our study and the data characteristic.

Discriminant analysis and logistic regression have widespread application to identify the group to which the object belongs, and to explain the base of object's group membership based on a set of independent variables (Hair et al., 2006). When assumptions are met, two-group discriminant analysis and logistic regression give comparable results. However, the former strictly relies on meeting the assumptions, including normality of independent variables (Hair et al., 2006), which is not the case for us. Logistic regression was chosen over this method also because it has straightforward tests, and there are a wide range of diagnostics available.

SEM, which is used in a number of related work (Alexander and Thomas, 2011; Espadanal and Oliveira, 2012; Gupta et al., 2013; Hsu, 2013; Wu, 2011), has an advantage over multiple regression in that it allows researchers to examine a number of dependence relationship simultaneously (Hair et al., 2006). However, logistic regression is preferable for several reasons. First of all, SEM assumes multi-normal distribution. Second, as we aimed to treat the independent variables as categorical, there would be too many dummy variables involved in the final model. The model would be therefore difficult to interpret, and this may affect conclusion validity. Third, the logistic model allows us to observe the change in odd-ratio when the degree of impact changes. Finally, SEM is most appropriate when a researcher uses multiple constructs, each represented by several measured variable (Hair et al., 2006). In our case, we used one measure for each construct.

### 5.3.1. Stage I – data distribution

The first step focused on understanding the distributions of the independent variables. We analyzed the level of concerns over the barriers using median, mode and data distribution. The shape of the distribution captured a general view of how the respondents feel about the relevance of the identified factors. We applied a *net stacked distribution graph* for visualization. The graph offers simple, yet comprehensive, visualization of numerous Likert-scaled variables. As a result, it is easier to capture and compare the skew between the total negative and positive responses among variables than a tradition approach, such as histogram.

### 5.3.2. Stage II – bivariate analysis

The second stage aimed at examining the relationship between each of the independent variables and the dependent variable, that is to understand to which extent each barrier is relevant for the adoption decision. This was done by checking the difference in the distributions of the concerns between the group of adopters and non-adopters. Nonparametric *Mann–Whitney–Wilcoxon* tests were applied to assess the equality of underlying distributions and to test the first set of hypotheses (*H_individual*). The null hypothesis states that the two distributions do not differ by a location shift. The nonparametric test is considered appropriate in our case due to the ordinal scale of the independent variables (Siegel and Castellan, 1988). The significance level was set at 0.05, meaning that if the null hypothesis is true, there is a 5% chance of incorrectly rejecting it. The rejection indicates a relationship between the barrier concern and the adoption decision.

*Mann–Whitney–Wilcoxon test.* The Mann–Whitney–Wilcoxon test (Sheskin, 2000) achieves a result comparable to the Mann–Whitney U test. The test checks whether distributions from two samples do not significantly differ by a location shift. It is used when the data is in rank order format or has been transformed into one from an ordinal/internal/ratio format. The test is often used when the assumptions of the *student t-test*, such as normality and homogeneity of variance assumption, are violated.

The Mann–Whitney–Wilcoxon test assumes that (1) each sample is selected randomly from the population; (2) two samples are independent; (3) the original variable is a continuous random variable; and (4) the shapes of the underlying distributions from which the samples are drawn are identical. Regarding the fourth assumption, a *Kolmogorov–Smirnov* test can be used for checking whether two samples drawn from each group can be considered identical in shape.

### 5.3.3. Stage III – multivariate analysis

The third stage focused on understanding the accumulative power of the major barriers in explaining variation in the enterprise cloud adoption decision. The analysis is classified as multivariate analysis because two or more independent variables are combined together to predict the value of a dependent variable (Meyers et al., 2006). In this case, logistic regression was applied to assess the discriminating power of the collections of factors in distinguishing a group of organizations that adopted the cloud from a group of organizations that did not.

At this level, further analysis is needed to identify the effect of contextual variables on the relationship between the barrier concerns and adoption decisions. Since the relationship we looked for might be partially confounded by the business domain and geographical differences, we asked *what is the discrimination power of the model when the contextual variables are controlled?* We applied the hierarchical regression method to evaluate the effects of contextual variables by entering the contextual variables into the first block of the regression model and the barriers into the second block.

The specification of the regression procedure needs to be clarified at this point. This includes the scale of the independent variables, the coding scheme, and the variable inclusion, before moving on to the regression model.

- Scale. There has been a long debate regarding the use of Likert-scaled variables in the regression analysis (Carifio and Perla, 2007). Likert assigns successive integers to ordered categories. In many cases researchers assume an equally-spaced difference between adjacent response categories and treat an ordinal Likert-scaled variable as an interval variable in the regression. In other cases, the Likert variable is treated as a categorical variable, thus using several dummy variables to represent its value in the regression model. In this analysis, we treat Likert-scaled variables as categorical variables. This is due to a lack of widely accepted indication of when it is legitimate to use Likert items as interval in regression analysis. Treating variables as categorical also allows us to observe changes in the odds-ratio as the degree of impact increases.
- Variable coding. Categorical variables need to be coded into a series of variables which then can be entered into the regression model. We used a simple coding scheme for independent variable encoding, as presented in Table 6. With this simple encoding, the category of "totally disagree", which was given the lowest score among the five scales, was treated as a reference category. Each category of the independent variable would be compared to the reference category during the analysis. Please note that the choice of coding scheme does not affect the power of the model.
- Variable inclusion. We applied the forward stepwise method to the regression to preserve the parsimonious property for the resulting model. With the forward variable selection, the process started with a constant-only model and the variables were entered one at a time into the model. The selection utilized the Likelihood Ratio statistic which tests the change in −2LL (two times the log of the ratio of the likelihood functions) between steps to determine which variable to add to the model and if previously included variables need to be removed. The probability for stepwise was set to 0.05 for the variable entry and 0.10 for variable removal.

**Table 6**
Categorial variables coding.

| Variable score | Parameter coding | | | |
| --- | --- | --- | --- | --- |
| | (1) | (2) | (3) | (4) |
| 1 (Totally disagree) | −0.200 | −0.200 | −0.200 | −0.200 |
| 2 (Disagree) | 0.800 | −0.200 | −0.200 | −0.200 |
| 3 (Neutral) | −0.200 | 0.800 | −0.200 | −0.200 |
| 4 (Agree) | −0.200 | −0.200 | 0.800 | −0.200 |
| 5 (Totally agree) | −0.200 | −0.200 | −0.200 | 0.800 |

*Logistic regression models.* The central concept underlying logistic regression is the logit, i.e. the natural logarithm of an odds ratio (Peng et al., 2002). The logistic model is well suited for describing and testing a hypothesis about relationships between a categorial outcome variable and one or more categorial or continuous predictor variables. The model has the form:

$$\text{logit}(Y) = \ln\frac{\pi}{1-\pi} = \alpha + \beta_1 X_1 + \beta_2 X_2 + \cdots + \beta_p X_p,$$

where $\pi$ is the probability of the event, $\alpha$ is the intercept, $\beta_i$ are regression coefficients and $X_i$ are predictors. $\alpha$ and $\beta_i$ are estimated by the maximum likelihood method. In other words, $\pi = \text{Probability}(Y = \text{outcome of interest} | X_1 = x_1, X_2 = x_2, \ldots, X_p = x_p)$ and therefore,

$$\pi = \frac{e^{\alpha+\beta_1 X_1+\beta_2 X_2+\cdots+\beta_p X_p}}{1+e^{\alpha+\beta_1 X_1+\beta_2 X_2+\cdots+\beta_p X_p}}$$

The null hypothesis states that all the coefficients are equal to zero. A rejection implies that at least one coefficient is not equal to zero. This means that the logistic regression equation predicts the probability of the outcome better than the mean of the dependent variable $Y$.

## 6. Results

We analyzed the central tendency and distribution of the variables and evaluated the hypotheses at the individual and integrated level to answer the research question. The results are presented in this section. It is followed by the reflection of the findings and implications for research and practitioners in Section 7.

### 6.1. Data distribution

Fig. 2 illustrates the respondents' concerns over six major barriers when making a decision on cloud adoption. The net stack graphs correspond to descriptive statistics of each variable presented in Table 7. Total responses for each barriers range from 355 to 380. Non-answers could be possibly due to uncertainty of the respondents about the impact of certain barriers on the decision in their particular cases.

The bar graphs are vertically ordered from the most relevant barriers to the least relevant one, based on a frequency count of the responses with a score four (agree) and five (totally agree). Availability, data privacy, and integration with current enterprise systems form a group of the most important factors during the decision making. 90.4% of the respondents considered it important or highly important to have close to 100% service uptime. 87.6% recognized a data privacy issue as influential or highly influential, and 79.4% agreed or highly agreed that the integration requirements impact their decision to adopt the cloud. Other factors perceived as important by more than half of the respondents are portability (67.9%) and security (60%). The least important barrier among the six appears to be migration complexity. Only 35.5% of the respondents considered migration complexity a major issue, while 39.5% seemed to accept the complexity of migration tasks, or believed that it could be done straightforwardly in their context. However, it should be noted that one fourth of the respondents have neutral opinion about this barrier. This could possibly suggest decision uncertainty due to company's insufficient assessment on the migration process and tasks involved.

**Table 7**
Descriptive statistics of the independent variables.

| Barriers | Percentile | | | Mode | # Responses |
|---|---|---|---|---|---|
| | 25 | 50 (Median) | 75 | | |
| Availability | 4 | 4 | 5 | 5 | 355 |
| Data privacy | 4 | 5 | 5 | 5 | 362 |
| Integration | 4 | 4 | 5 | 4 | 379 |
| Portability | 3 | 4 | 5 | 4 | 380 |
| Security | 3 | 4 | 4 | 4 | 360 |
| Migration | 2 | 3 | 4 | 2 | 380 |

### 6.2. Results from bivariate analysis

The boxplot comparisons of the two groups, i.e. adopters and non-adopters, regarding the influence of barrier concerns on the decision to adopt cloud is presented in Fig. 3. Availability, data privacy and integration issues appear as influential or highly influential for both groups. However, when considering the security, its distribution shifts toward a higher level of influence for the group of non-adopters. The results of two-sample nonparametric tests are summarized in Table 8. A *Kolmogorov–Smirnov* (KS) test was applied to check the similarity in shape of two underlying distributions, which is one of the prerequisite to perform a *Mann–Whitney–Wilcoxon* (Wilcoxon) test. The results from KS indicate the significant difference in the distribution of adopters' and non-adopters' perceptions on security and migration complexity ($p$-value is 0.000 and 0.045, accordingly). The Wilcoxon's test results suggest a difference in location for the compared distributions in the case of data privacy ($p$-value is 0.044), portability ($p$-value is 0.047), and security ($p$-value is 0.000). This suggests that non-adopters tend to have more concerns about these three issues than adopters. The significant differences provide initial evidence of a relationship between the barrier concern and the likelihood of adopting cloud services. To conclude, the null hypotheses for the bivariate analysis ($H_0\_individual$) are rejected in three cases. There is sufficient evidence to indicate a location shift to a higher level of concern with regards to data privacy, portability, and security in a group of non-adopters.

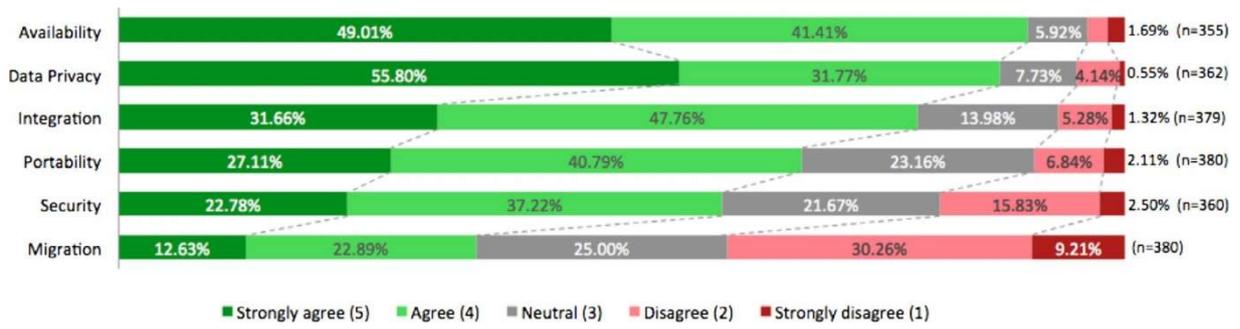

**Fig. 2.** Net stacked distribution of the concerns over six major barriers

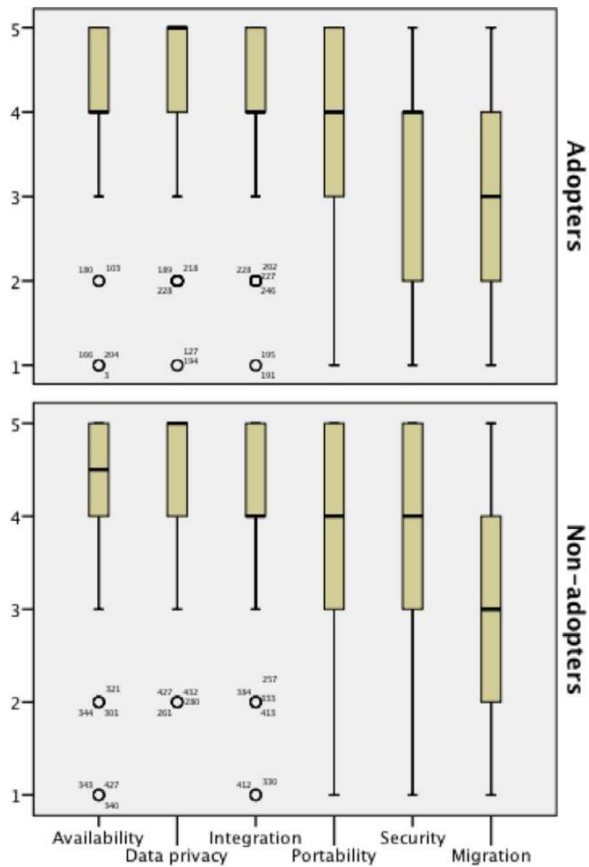

**Fig. 3.** Boxplot showing the distributions of barrier concerns between the group of adopters and the group of non-adopters *(the bottom and the top of the box represent the first and the third quartile, and the band inside the box is the median).*

### 6.3. Results from multivariate analysis

A logistic model was fitted to the data to evaluate the research hypothesis at the integrated level. The control variables were entered into the model in the first block as a binary dummy variable. Afterward, we applied the stepwise (forward) method as a criteria for including a predictor in a model. The significance of the model was evaluated and the importance of each variable was assessed. The baseline model contains only the constant, leading to the prediction of the more common cases. In our case, more samples (57.4%) are cloud adopter. Security (SE) and interoperability with an existing system (IN) were found to significantly increase the accuracy of the model and were entered into the model in the first and the second step, respectively. An inclusion of other variables did not significantly improve the model and thus they were dropped from further analysis. We also checked for the interaction among the predictors. However, the estimated coefficient for the interaction term was not statistically significant, and therefore not included in the final model. In this section, we first outline the results of model evaluation, and the interpretation of model is given afterward.

#### 6.3.1. Model evaluation
Evaluations of the final logistic regression model include the overall model evaluation, statistical tests of individual variables, goodness-of-fit statistics, and validations of predicted probabilities.

##### 6.3.1.1. Overall model evaluation.
A logistic model provides a better fit to the data if it demonstrates an improvement over the baseline model. The omnibus test of model coefficients given by SPSS contains the model chi-square ($\chi^2$), a statistical test of the null hypothesis that all the coefficients are zero. The test gives a $\chi^2$ *(degree of freedom = 14)* of 57.121. The null hypothesis ($H_{o\_integrated}$), stating that the contribution of independent variables included in the research framework in explaining the variation in the cloud adoption decision is zero, is rejected because the significance is less than 0.05. Therefore, we conclude that the set of independent variables improves prediction of the outcome.

##### 6.3.1.2. Statistical test for individual predictor.
The statistical significance of individual coefficients is tested using a Wald chi-square statistics. On the basis of the stepwise regression, security and interoperability between the cloud and an existing system are statistically significant predictors at the 0.05 level (*p*-value is 0.000 and 0.047, respectively) (Table 9).

##### 6.3.1.3. Goodness-of-fit statistics.
The results for the model fit are summarized in Table 10. The test assesses the fit of the predictions and the actual outcomes, i.e., whether an organization proceeds with cloud adoption. The Hosmer and Lemeshow test yields a $\chi^2(8)$ of 9.481 and a significance of 0.303. The nonsignificant *p* value suggests that the null hypothesis, which states that the prediction and outcomes do not differ, is not rejected. This result indicates an acceptable match between the predicted and actual values.

The power of the model in explaining data variation is also assessed by −2 log likelihood (−2LL) and pseudo $R^2$. The pseudo $R^2$ is not statistically equivalent to $R^2$ produced in ordinary least squares regression, as the logistic models are maximum likelihood estimates. Low pseudo $R^2$ values in logistic regression are the norm (Hosmer and Lemeshow, 2000). −2LL is useful when comparing the fit between two models. The model yields a −2LL of 423.144. The Cox and Snell pseudo $R^2$ is 0.150, and the Nagelkerke pseudo $R^2$ is 0.201. Usually the Nagelkerke pseudo $R^2$ is preferable as it achieves a maximum value of 1.

#### 6.3.2. Discriminating power
To evaluate the discriminating power of the final model, we compare it against the baseline, zero-predictor model. The classification table (Table 11) summarizes correct and incorrect predictions, where the columns correspond to the model predicted value and the rows correspond to the actual cloud adoption decision. As mentioned, the baseline model always predicts the adoption of cloud computing, due to this being the more common case. This results in a prediction rate of 57.4%. The final model with two predictors yields a correct prediction rate of 79.2% for the adopter group and of 57.3% for the non-adopter group. This leads to overall predictive accuracy of 69.9%.

**Table 8**
Nonparametric tests of the distributions of barrier concerns between a groups of adopters and non-adopters.

| Barrier concern | Number | | Sig. (*p*-level) | Mean rank | | Sig. (*p*-level) |
|---|---|---|---|---|---|---|
| | Adopters | Non-adopters | KS | Adopter | Non-adopter | Wilcoxon (1-tailed) |
| Availability | 203 | 152 | 0.977 | 177.49 | 178.68 | 0.453 |
| Data privacy | 206 | 156 | 0.179 | 174.25 | 191.08 | 0.044* |
| Integration | 214 | 165 | 0.378 | 194.19 | 184.57 | 0.180 |
| Portability | 216 | 164 | 0.210 | 182.69 | 200.79 | 0.047* |
| Security | 205 | 155 | 0.000** | 160.20 | 207.35 | 0.000** |
| Migration | 216 | 164 | 0.045* | 184.32 | 198.63 | 0.098 |

**Table 9**
Relevant statistics for interpreting logistic regression results.

| Statistic | Explanation |
|---|---|
| −2 log likelihood (−2LL) | A measure of model estimation fit. The minimum value for −2LL is 0, which represents a perfect fit. The lower the −2LL, the better fitting the model. |
| Cox and Snell $R^2$ | The pseudo $R^2$ is interpreted in a manner similar to $R^2$ of multiple regression. It reflects the amount of variation accounted for by the model, with a maximum value of 1 indicating a perfect fit. This particular measure gives a maximum of less than 1. |
| Nagelkerke $R^2$ | A modification of Cox and Snell $R^2$ that has a value range of 0–1. |
| Hosmer–Lemeshow goodness-of-fit | This test provides a measure of predictive accuracy, based on the actual prediction of the dependent variable. It groups cases into $n$ groups, then compares the number of actual and predicted values using chi-square statistic. |
| Wald statistic | The test to assess significance of each coefficient, that is to see if it is significantly different from 0. If the logistic coefficient is statistically significant, then we can interpret the impact of that independent variable on the estimate probability. |
| Odds ratio | The ratio of the odds of the reference and compared groups. It represents how likely it is for the event to occur with presence of a certain condition. An odds ratio of 1 indicates that the event is equally likely in the reference and the compared group. When the odds ratio is greater than 1, the probability of the event occurring is *higher* with unit increase in the independent variable. An odds ratio less than 1 indicates that the probability of the event occurring is *lower* with unit increase in the independent variable. |

**Table 10**
Results for model fit.

| −2 log likelihood | Cox and Snell $R^2$ | Nagelkerke $R^2$ | Hosmer–Lemeshow goodness-of-fit |
|---|---|---|---|
| 423.144 | 0.150 | 0.201 | 0.303 |

**Table 11**
Classification table.

| Actual | | Predicted | | |
|---|---|---|---|---|
| | | Decision | | Percentage correct |
| | | Adopt | Not adopt | |
| Baseline | Adopt | 202 | 0 | 100.0 |
| | Not adopt | 150 | 0 | 0.0 |
| | Overall percentage | | | 57.4 |
| Final | Adopt | 160 | 42 | 79.2 |
| | Not adopt | 64 | 86 | 57.3 |
| | Overall percentage | | | 69.9 |

### 6.3.3. Interpreting the coefficients and odds ratio

Table 12 summarizes the results of the final model evaluation as well as the coefficients, odds ratio, and test statistics of the variables included in the final model.

A correlation coefficient ($\beta$) indicates the amount of change in the log odds ratio when there is a one-unit change in the predictor variable with other variables in the model held constant. Therefore $\exp(\beta)$ is the odds ratio, representing the ratio of the odds for the adopter and non-adopter group. An odds ratio of 1 indicates that the event (i.e. the decision was taken not to adopt the cloud) is equally likely in the reference and the compared group. An odds ratio greater than 1 suggests a higher likelihood that the compared group is a non-adopter. An odds ratio less than 1 indicates a lower likelihood that the compared group is a non-adopter.

According to the variable coding (Table 6), a *strongly disagree* level of agreement is used as a reference category, and each level is compared to the reference level. The odds ratios for the security concerns are approximately 4.5, 14.9, 17.4 and 26.4, from the level of *disagreement* to the level of *strong agreement*. The continuous increase of the values indicates that when we have statistically controlled for interoperability and contextual differences, the more the security issue is concerned, the higher the likelihood that organizations would decide not to adopt cloud services. So the influence of security issues is strong. When one is highly concerned about sharing security responsibility with a cloud provider, the likelihood of deciding not to adopt cloud increases by an order of 26.4, as compared to the case that this condition is not relevant or concerned.

Difficulties involving compatibility between cloud and in-house systems do not appear to prevent firms to adopt cloud services. The integration (IN) regression coefficient suggests negative relationship between the degree of concerns and non-adoption tendency. This implies that the adopters are aware of the necessity to promote interoperability between the systems and have very likely searched for a solution in this area.

Regarding the business domain, it appears that firms operating in the information and communication technology (ICT) have more propensity to adopt cloud. This result is suggested by significant coefficient ($p < 0.05$) and the value of odds ratios (<1.00) of the first control binary variable in Table 12. Geographical differences do not appear to impact the adoption tendency.

## 7. Discussion and implications

### 7.1. Discussion of the findings

We sought to understand the relevance of major technical barriers and the decision to adopt the cloud in organizations. Our review of industry cloud surveys reveals a wide range of barriers and limitations that might hinder mainstream adoption of cloud services. Compared to the collection of barriers gathered in the previous work, such as Khajeh-Hosseini et al. (2011), the specific set of barriers presented in Table 2 is considered the most relevant subset of cloud adoption barriers for industrial contexts. Our survey confirms the findings of existing surveys (such as Everest Global (2013); IDG Enterprise (2013); NTT Europe Ltd. (2013) conducted in 2013) that majority of (prospect) adopters are conscious of potential pitfalls on service availability, data privacy, integration, portability and security. Their relevance in contexts were judged significant by 90%, 88%, 80%, 68% and 60% of the respondents, respectively. Migration complexity appeared to be the least important barrier based on the frequency count, only 35% of the respondents considered it a major issue.

Focusing particularly on the major technical barriers, our analysis suggests that not the entire set of barriers significantly influence the adoption decision, even though they are consistently assessed during the decision making process. Table 13 summarizes the findings on barriers, that distinguishes potential adoption inhibitors from other relevant barriers that may not play an insurmountable role in an early adoption stage.

Potential adoption inhibitors were identified when we compared a subgroup of adopters and non-adopters. Wilcoxon test suggests a shift of nonadopter's distribution toward a higher level of concerns when considering security, data privacy, and portability. Furthermore, the logistic model suggested security as the most critical factor that indicates the cases of non-adoption in our sample. There is an up to 26-fold increase in the likelihood of non-adoption if a provider is perceived not to be able to provide adequate security measures. The open-end comments reveal various security and data privacy concerns that trigger non-adoption, including access mechanisms, shared

**Table 12**
Variables in the equation (final model).

| Variable | B | SE | Wald | df | Sig. | Odds ratio | 95% CI for odds ratio | |
|---|---|---|---|---|---|---|---|---|
| | | | | | | | Lower | Upper |
| SE | | | 20.211 | 4 | 0.000 | | | |
| SE(1) | 1.493 | 1.403 | 1.133 | 1 | 0.287 | 4.451 | 0.285 | 69.560 |
| SE(2) | 2.699 | 1.405 | 3.688 | 1 | 0.055 | 14.859 | 0.946 | 233.428 |
| SE(3) | 2.855 | 1.400 | 4.156 | 1 | 0.041 | 17.366 | 1.116 | 270.145 |
| SE(4) | 3.273 | 1.399 | 5.471 | 1 | 0.019 | 26.395 | 1.700 | 409.870 |
| IN | | | 9.621 | 4 | 0.047 | | | |
| IN(1) | −2.522 | 1.443 | 3.057 | 1 | 0.080 | 0.080 | 0.005 | 1.357 |
| IN(2) | −1.139 | 1.354 | 0.707 | 1 | 0.400 | 0.320 | 0.023 | 4.551 |
| IN(3) | −1.750 | 1.332 | 1.726 | 1 | 0.189 | 0.174 | 0.013 | 2.365 |
| IN(4) | −2.145 | 1.333 | 2.588 | 1 | 0.108 | 0.177 | 0.009 | 1.597 |
| Control binary variable | | | | | | | | |
| Firms operating in ICT domain | −0.949 | 0.246 | 14.883 | 1 | 0.000 | 0.387 | 0.239 | 0.627 |
| Firms operating in Africa | 1.600 | 1.245 | 1.653 | 1 | 0.199 | 4.955 | 0.432 | 56.841 |
| Firms operating in Asia | 0.055 | 0.963 | 0.003 | 1 | 0.954 | 1.057 | 0.160 | 6.982 |
| Firms operating in Australia | 0.726 | 1.091 | 0.443 | 1 | 0.506 | 2.067 | 0.243 | 17.549 |
| Firms operating in Europe | 0.746 | 0.932 | 0.641 | 1 | 0.423 | 2.109 | 0.339 | 13.101 |
| Firms operating in North America | 0.461 | 0.929 | 0.246 | 1 | 0.620 | 1.586 | 0.257 | 9.794 |
| Constant | −0.649 | 0.941 | 0.475 | 1 | 0.491 | 0.523 | | |
| Test | | | $\chi^2$ | df | Sig. | | | |
| Overall model evaluation | | | | | | | | |
| Omnibus test of model coefficients | | | 57.121 | 14 | 0.000 | | | |
| Goodness-of-fit test | | | | | | | | |
| Hosmer and Lemeshow test | | | 9.481 | 8 | 0.303 | | | |

**Table 13**
Summary of results

| Barrier | Relevance[a] | Potential inhibitor[b] |
|---|---|---|
| Availability | | |
| Data privacy | | |
| Integration | | |
| Portability | | |
| Security | | |
| Migration | | |

[a] indicates that majority of the respondents find the barrier important for the decision to adopt the cloud (based on frequency count).
[b] indicates a shift of distribution to a higher level of concern for a group of non-adopters (based on Wilcoxon).

network and infrastructure, compliance to HIPAA, and protection for protected health information (PHI). One non-adopter suggested that cloud providers *"must show bottom-line advantages and solid security"*

The significant impact of security risk was also captured in a more specific scenario (Alexander and Thomas, 2011; Gupta et al., 2013; Lian et al., 2014). Alexander and Thomas (2011) found that among the major risk factors driving SaaS adoption intention, security risks were the dominant factor, more so than performance and economic risks. Security concerns were perceived as the most critical concerns in the healthcare context (Lian et al., 2014). Gupta et al. (2013) found that better security also leads to an increase of adoption intensity.

Data privacy is another critical concern, even thought its negative effect seems less strong than security, as suggested by Wilcoxon and the stepwise method. As for financial industry, private cloud remains a preferring deployment choice for this reason (Wenge et al., 2014). Business cases that involve Personal Identifiable Information (PII) and other confidential data are considered ineligible for the cloud. Financial institutes acknowledge the legal instruments to mitigate data privacy risks for their customers. Nevertheless, a recent study conducted in a healthcare domain reveals that trust in the cloud provider, technology advancement, and privacy-preserving regulatory can significantly mitigate the privacy concern (Ermakova et al., 2014).

Cloud computing raises the lock-in or portability issue to a profound level, and the providers obviously gain certain advantages from keeping customers locked-in to their environment. This is a great concern for consumers from the early adoption stage, to be able to bring their asset back on premise or to move to another provider if needed.

The result from Wilcoxon demonstates this criticality. Nevertheless, it was not always considered as a primary technical concern in exisiting quantitive studies, such as Alexander and Thomas (2011); Lumsden and Gutierrez (2013); Tehrani (2013).

Service availability, integration and migration complexity were not determined as potential adoption inhibitors. As for availability, its low influence is somewhat surprising. However, it can be partially explained by the readiness of fundamental network infrastructure in our era, resulting in sufficient reliability and performance to support normal operation of the cloud. This finding is partly consistent with the results achieved by Alexander and Thomas (2011) as they examined the SaaS non-adopter sample and found that only financial and security risks have a strong and negative impact on the adoption intension. Factors such as speed, network reliability, interoperability, as well as, other managerial and strategic risks are not IT executives' dominant concern.

The low inhibiting potential of migration complexity and integration concerns may indicate willingness of enterprises for tasks required. In fact, based on the regression results, realizing the challenges of system and data integration does not seem to stop the cloud initiative. Although, complexity of migrating legacy applications was mentioned as challenging several times in the comment section. Different aspects of complexity were investigated in previous studies (Espadanal and Oliveira, 2012; Lian et al., 2014; Lumsden and Gutierrez, 2013; Tehrani, 2013), yet obtaining inconclusive results. In Tehrani (2013), perceived complexity to use the cloud does not appear to have a significant negative impact on the adoption likelihood. Analysis of cloud adoption in UK also indicates it as the least important factors (Lumsden and Gutierrez, 2013). In contrast, usage complexity was found to be a significant inhibitor for manufacturing and service firms. The importance of complexity was ranked quite high (5 out of 12) in the study conducted in a healthcare domain, as this work considers a more comprehensive view of complexity, including complexity to migrate, develop, maintain, and learn to use the cloud services.

Considering the studies that investigate cloud adoption from a more holistic approach (Table 3), the technological context has been identified as an influential aspect for cloud adoption (Hsu, 2013; Lian et al., 2014; Lumsden and Gutierrez, 2013). Nevertheless, technical barriers, although playing an important role, are not always an insurmountable hurdle. Lumsden and Gutierrez (2013) suggest that once organizations identify the feasibility of adopting cloud services, the benefits outweigh the perceived complexity. The relative advantage of the technology was also found to have a stronger influence on the adoption than the barriers (Alexander and Thomas, 2011; Gupta et al., 2013; Hsu, 2013; Wu, 2011).

### 7.2. Theoretical contributions

Firstly, the empirical result has provided support that our model based on major technical barriers is capable of distinguishing the case of cloud adoption and non-adoption. Previous studies on determinants for enterprise cloud adoption employed a multi-dimensional framework, resulting in a model with a relatively higher explanation power. However, our simple model based on one dimension of the technology aspect has led to approximately 70% predictive accuracy, demonstrating the importance of this particular angle on the growth of cloud adoption. The second contribution is that it advances our understanding about the importance of each technical concern on the decision making process. Concerns about security is the dominant factor for overall risk perception.

### 7.3. Implications for practitioners

The study provides a comprehensive view on cloud computing adoption barriers, especially the ones relevant to the majority of enterprise adoption. It also reveals the differences between adopters' and non-adopters' perceptions toward the barriers. Several implications for cloud providers, and (prospect) consumers cloud be derived from the findings. For the providers, the reviews of the 2011–2013 industrial surveys and our survey identify the market's consistent concerns over a similar set of barriers. This implies unclarity or a lack of mechanisms for risk assessment that is applicable in large contexts. Based on the analysis, prospect adopters might easily turn down adoption consideration by an ambiguous security perception of the provider. Often the security concerns are boiled down to a matter of trust, which is built over time. Therefore, it is advisable for the provider to be certified to the respected security standards to attract newcomers. Demonstrating the security controls, in comparison with competitors, might help accelerate the adoption decision. In our view, the high security odds ratio also implies high provider switching cost, especially in the critical contexts. This creates an opportunity for the provider to encourage more spending from their existing consumers through positive experiences. For consumers, we suggest applying a holistic and measurable approach to assess and understand the effectiveness of the cloud security controls. Provider transparency is essential to identify the gap where the security and reliability objectives are not met, so that the corrective actions can be effectively taken.

## 8. Threats to validity

A fundamental question concerning the results is how valid the results actually are. This section discusses what we consider to be the most important validity issues in this study based on four categories suggested by Wohlin et al. (2012).

### 8.1. Construct validity

Construct validity is concerned with the relation between the theory and the observation. We must ensure generalizability of the results to the underlying concept. There are three distinct issues, related to the measurement instrument, the design of constructs, and the unit of analysis.

Regarding the *measurement instrument*, there is always a risk that the question is misunderstood. This could be caused by insufficient background knowledge of a respondent about the related concepts and ambiguity of the instrument itself. In this context, an additional risk is unclarity over the term *cloud computing* and other technical terms, such as vendor lock-in. Main actions taken to reduce the mentioned risks include: (1) presenting a definition of cloud computing and examples of different types of cloud services at the beginning of the survey, (2) avoiding the use of technical terms by replacing them with a simple phrase that explains the underlying concept, (3) following the guideline given by Oppenheim (1992) when designing the survey and formulating questions, (4) pre-testing the instrument with industrial professionals, to assess if the questions are understood as intended and to determine the time required to complete the survey, and (5) allowing respondents to skip a question when they are uncertain about the involved concept.

There are two limitations of the current study regarding the design of constructs and an ability of the selected measures to reflect a construct to be studied. First of all, *mono-operation bias* occurs when a study includes a single independent variable, case, subject or treatment. In our case the use of a single construct may under-represent a multi-facet nature of each barrier. For example, a respondent may consider availability from an aspect of service uptime, and/or, sufficient internet bandwidth; security concerns are considered from an aspect of identity control, infrastructure management, and/or, trust. Measuring the concept using multiple constructs and considering an extended set of barriers to see how the prediction accuracy improve are suggested as future work.

The second design limitation is concerned with *confounding constructs and levels of constructs* (Wohlin et al., 2012). In some cases, it is not the presence of the construct, but the level of construct affecting the outcome (Wohlin et al., 2012). In our case, the adoption decision is measured using a binary response representing a case of adoption and non-adoption. However, the adoption decision could have been designed as a multiple-stage variable, including a case that a company has not considered the adoption and a case that the company is considering but has not decided. A more fine-grained measure could have impacted an inclusion of cases for the analysis.

The last issue is about a unit of analysis. The ideal case to reflect relevancy between barriers and an adoption decision is to obtain one response from each sampling organization made by a key decision maker. However, this cloud not be achieved by conducting a self-selection survey. We have handled this limitation by considering professional's opinion (rather than organization) as a unit of analysis.

### 8.2. Internal validity

Internal validity concerns the causal relationship between the dependent and independent variables. If the causality is observed, we must be certain that it is not a result of factors not being controlled or measured. Possible *confounding factor* in this context is the role of a respondent in the cloud community, and the purpose for which the cloud is used. Firstly, the previous research suggests that cloud consumers, providers, advisory and IT decision makers may have different opinions about the relevance of barriers in adoption decision (KPMG International, 2012b). If this information had been collected in our work, the analysis could have been extended to uncover its effect on the outcome. Secondly, companies may adopt the cloud for internal use, or leverage public cloud services to build another cloud solution for their consumers. The latter is one of the major trend in Software Engineering and software startup community. The six barriers included in the study should be relevant for both cases. However, the degree of importance could be slightly different, for example interoperability might play a more important role in the latter case.

### 8.3. Conclusion validity

Conclusion validity is concerned with an ability to derive legitimate conclusions from observations. A selection of statistical models and model configurations directly affect conclusion validity. We adopted a nonparametric test, i.e., Mann–Whitney–Wilcoxon, to evaluate the hypotheses at the individual level, as the normality assumption of the parametric test is violated. At the integrated level, we chose to treat Likert variables as categorial, which allows us to observe the change in the odds-ratio at a particular scale. The variable was included in the regression model based on a forward stepwise method, to avoid multicollinearity.

*8.4. External validity*

The external validity is concerned with generalizability. The conditions that could limit an ability to generalize the result outside the scope of the study include three types of interaction with treatment: selection, setting, and time (Wohlin et al., 2012). In terms of *selection*, given that the survey was posted in online communities, we have no control over the samples. A control, such as a response rate, is therefore not applicable. Larger sample size may contribute positively to the diversity, and thus generalizability of the survey finding. We argue that our survey has received an acceptable number of responses in comparison to other international surveys (Table 1) to draw a valid conclusion (443 in total, and 352 complete responses). This number is also acceptable comparing with other self-selection surveys in software engineering, such as surveys on agile practice (Barabino et al., 2014; Kurapati et al., 2012) and software testing (Kanij et al., 2013). Apart from the size, the generalizability is affected by coverage bias. As our population of interest is organizations that have considered adopting cloud services to serve their organizational objectives, the bias is partially reduced by (1) placing the survey invitation on many cloud computing discussion groups whose target respondents fit our research aim, and (2) checking the professional profile of group members, which was possible in most cases (especially in LinkedIn's discussion groups). Additionally, the sample demographics is quite diverse, as shown in Table 5. Even though the majority of the sample is from the ICT domain, it corresponds to other large-scaled surveys.

Another related threat is *time interaction*, the importance of barriers might become irrelevant in time as the technology and supporting mechanisms reach a higher level of maturity. We argue, based on the literature survey, that the findings of our work are still relevant in the cloud computing community. Large concerns over several issues, including security, data privacy and integration, remain and have been captured in recent surveys.

## 9. Conclusion

Technical barriers and security-related risks may hold back cloud adoption in organizations. As the value of the cloud is becoming more evident through industrial success stories, the real question is why certain enterprises are still reluctant to migrate their IT resources into the cloud. The first step to address this question, with a goal of reducing the barrier to entry, is to identify the concrete factors that influence the adoption decision. Our study focuses on six major barriers, namely availability, portability, integration, migration complexity, data privacy, and security. Six hypotheses regarding the relevance of each barrier and the adoption decision, as well as a hypothesis at the integrated level, were evaluated. Three individual-level hypotheses were rejected. The results suggest significant increase in a level of concern on security, privacy, and portability (vendor lock-in) for non-adopter sample. The integrated-level hypothesis was also rejected, indicating that combined influence of barriers can be used to determine the case of non-adoption. The criticality of security was insisted by the logistic model. The other barriers (service availability, integration, and migration complexity) are not likely to stop a cloud adoption initiative. We conclude here by answering the posted research question that the major technical barriers yield about 70% accuracy for predicting the decision of cloud adoption in enterprises.

Future work should extend the scope of the current study by (1) investigating effects of the respondent's role and the purpose for which the cloud service is used on barrier perception, (2) considering an extended set of barriers to see how the prediction accuracy improves, and (3) integrating the statistical findings with a qualitative analysis to gain deeper variance explanation across business domains. In addition, on the basis of related work on cloud adoption determinants, meta-analysis can be conducted to combine results from different studies.

**Nattakarn Phaphoom** obtained a PhD from Free University of Bozen-Bolzano, and double master's degree in software engineering from Blekinge Institute of Technology, Sweden, and Free University of Bolzano-Bozen, Italy. Prior to studying the master's degree, she worked for 3 years for IBM Solutions Delivery Co., Ltd., Thailand. Her research interests include emerging technologies, innovation adoption, cloud computing, data analytics, and agile methods.

**Xiaofeng Wang** is a researcher at the Free University of Bozen-Bolzano. Her research areas include software development process, methods, agile software development, and complex adaptive systems theory. Contact her at xiaofeng.wang@unibz.it.

**Sarah Samuel** achieved a BSc in Information Systems and Management from Birkbeck, University of London. She has been working in technology since 2001 with a focus on Internet and telecommunications and has extensive experience in deploying network infrastructure on a global scale. Sarah is now employed by Astreya providing services to Google in Network Operations on projects across EMEA. She is originally from Jamaica, and has since lived in New York City, Barbados, Virginia, Winnipeg, Zimbabwe, Montreal, and now settled in London, UK. Her interests include cloud computing, artificial intelligence, scuba diving, figure skating, and Wing Tsun Kung Fu.

**Sven Helmer** is an Associate Professor at the Faculty of Computer Science at the Free University of Bozen-Bolzano, Italy. He obtained a PhD from the University of Mannheim, Germany, and an MSc in Computer Science from the University of Karlsruhe, Germany. His research interests include non-relational database systems, query optimization, index structures, route planning as well as interdisciplinary research in the areas of information systems and ethnography.

**Pekka Abrahamsson** is a full professor of computer science at Free University of Bozen-Bolzano. His research interests are centered on empirical software engineering, agile and lean software development and cloud computing. Contact him at pekka.abrahamsson@unibz.it.